\documentclass[aps,twocolumn,prd,showpacs,showkeys,preprintnumbers,superscriptaddress,nobibnotes,floatfix,longbibliography,nofootinbib]{revtex4-1}

\usepackage[T1]{fontenc} 
\usepackage[compat=1.1.0]{tikz-feynman}
\usepackage{pifont}
\usepackage[mathscr]{eucal}
\usepackage{multirow}
\usepackage{graphicx}
\usepackage{amsfonts}
\usepackage{amsmath}
\usepackage{amssymb}
\usepackage{epsfig}
\usepackage{slashed}
\usepackage{dcolumn}%
\usepackage{ulem}
\usepackage{color}
\usepackage{float}
\usepackage{placeins}
\usepackage{booktabs}
\usepackage[english]{babel}
\usepackage{microtype}
\usepackage{pifont}
\definecolor{myred}{RGB}{255, 0, 0}
\definecolor{myblue}{RGB}{0, 0, 255}
\definecolor{mygreen}{RGB}{0, 128, 0}

\usepackage[colorlinks,
linkcolor=blue,
citecolor=blue,
urlcolor=blue
]{hyperref}

\begin{document}
\title{Photon-Mediated Axino-Bino Scattering Process for the
LZ High-Energy Recoil Event}

\author{Xiangwei Yin}
\email{yinxiangwei@htu.edu.cn}
\affiliation{School of Physics, Henan Normal University, Xinxiang 453007, P. R. China}

\author{Quanyu Wang}
\email{wangquanyu@htu.edu.cn (corresponding author)}
\affiliation{School of Physics, Henan Normal University, Xinxiang 453007, P. R. China}

\author{Tianjun Li}
\email{tli@itp.ac.cn}
\affiliation{School of Physics, Henan Normal University, Xinxiang 453007, P. R. China}

\begin{abstract}
	
We for the first time investigate the photon-mediated endothermic axino-bino scattering process to explain the LUX-ZEPLIN (LZ) high-energy recoil event. We calculate the signals from non-relativistic halo and boosted dark matter (DM), including natural xenon isotopes, detector efficiency, and energy resolution. For a fixed Standard Halo Model (SHM) distribution, a simplified three-bin Poisson analysis yields a best-fit benchmark with mass $m\simeq506\,\mathrm{GeV}$ and mass splitting $\delta\simeq326\,\mathrm{keV}$ at the adopted upper coupling bound $g_\gamma=10^{-3}\,\mathrm{GeV}^{-1}$. It predicts approximately one accepted event, with $70\%$ in the middle bin. A boosted benchmark with $m\simeq82.87\,\mathrm{GeV}$ and $\delta\simeq0.626\,\mathrm{MeV}$ provides a comparable fit. Broader Gaussian incident energy distributions allow fitted masses down to a few GeV. Thus, both the halo and boosted DM scenarios can fit the LZ signal at different DM mass scales.

\end{abstract}
\maketitle


\section{Introduction}\label{Sec1}

The particle nature of dark matter (DM) remains unknown despite compelling gravitational evidence for its existence~\cite{Bertone:2004pz,Bertone:2016nfn,Planck:2018vyg}. Many direct detection experiments probe DM-nucleus interactions by searching for the nuclear recoils produced when DM scatters off detector nuclei~\cite{XENON:2018voc,LZ:2022lsv,XENON:2023cxc}. Recently, the LUX-ZEPLIN (LZ) experiment extended its recoil-energy search to approximately $270\,\mathrm{keV}$ and reported a candidate event at $248\pm23\,\mathrm{(stat)}\pm23\,\mathrm{(sys)}\,\mathrm{keV}$ in an exposure of $2.84$ tonne-years~\cite{LZ:2026axp}. The observation is in tension with the background-only hypothesis at a maximum local significance of $3.4\sigma$, reduced to a global significance of $2.6\sigma$ after accounting for the look-elsewhere effect. Although this is insufficient to claim a discovery, it motivates investigating whether DM scattering can produce high energy recoils without an accompanying low energy excess. Standard spin-independent elastic scattering with a momentum-independent DM-nucleon amplitude favors low energy recoils, so increasing the cross section to enhance the high energy recoil rate also raises the low energy rate~\cite{LZ:2022lsv,XENON:2023cxc,Lewin:1995rx,Jungman:1995df,Freese:2012xd,MarrodanUndagoitia:2015veg}. Exothermic down-scattering of an excited, heavier DM state can convert the mass-splitting energy into nuclear recoil energy, potentially making detectable recoils possible even when the incident DM kinetic energy is small~\cite{Graham:2010ca,He:2024hkr}. However, if the excited state is unstable, it may decay before the scattering in detector~\cite{Bell:2021xff}. Endothermic inelastic scattering offers a possible mechanism. The transition to a heavier DM state introduces a velocity threshold and can suppress low energy recoils~\cite{Smith:2001aa,XENON:2024iDM}. Recent studies have explored several interpretations of the LZ candidate. These include supersymmetric and electroweak DM scenarios, such as Higgsino, sneutrino, and bino-like models~\cite{Yin:2026jnn,Fan:2026kxx,Freese:2026sga,Visinelli:2026kgt,Du:2026guj,Lian:2026hpm,Roy:2026wzb}, as well as scalar or fermion DM models~\cite{Wang:2026ytg,Bandyopadhyay:2026gjw,Borah:2026zwf}. Other studies consider endothermic or exothermic inelastic scattering, including dark-photon models~\cite{Yamashita:2026ump,Zhu:2026dag,Tang:2026lgu}, axion- or ALP-portal models~\cite{Unwin:2026rdp,Yuan:2026djt,An:2026pkc}, and scenarios involving xenon excitation~\cite{Gu:2026vto}. Further analyses examine exothermic scattering and sideband constraints~\cite{Baer:2026fpy,Dent:2026bji,Xing:2026civ}, other inelastic-DM frameworks~\cite{Fan:2026hzw,He:2026idw,Ge:2026xax,Su:2026rwz,Das:2026uyy,Palmisano:2026kuj}, and DM absorption~\cite{Lou:2026idn} as well as boosted DM scenarios~\cite{Liang:2026coz,Kannike:2026qyl,Alhazmi:2026efz,Heikinheimo:2026kwp,Mahapatra:2026glu,Chauhan:2026udz}.

In this work, we present the first investigation of  photon-mediated axino–bino transition to explain the LZ high-energy recoil event. We consider the endothermic inelastic process $\widetilde a A\to\widetilde B A$, in which axino DM
converts into a heavier bino-like neutralino through the nuclear electromagnetic current~\cite{Covi:1999ty,Covi:2001nw}. We study both non-relativistic halo DM and boosted DM. Starting from the effective axino--gaugino interaction, we derive the non-relativistic differential cross section and employ the exact cross section and two-body kinematics for the boosted case. We calculate the signal for natural xenon, include detector efficiency, energy resolution, and exposure, and fit the LZ data using a simplified three-bin Poisson likelihood.

For non-relativistic DM, we scan the mass $m$, positive mass splitting $\delta$, and transition coupling $g_\gamma$.
We integrate the scattering rate over the laboratory-frame Standard Halo Model (SHM) velocity distribution~\cite{Drukier:1986tm}, accounting for the speed-dependent flux. The best-fit benchmark has
$m\simeq506\,\mathrm{GeV}$, $\delta\simeq326\,\mathrm{keV}$, and $g_\gamma=10^{-3}\,\mathrm{GeV}^{-1}$, at the adopted upper coupling bound. It probes the high-speed halo tail and predicts approximately one accepted event, with about $70\%$ in the middle bin.

For boosted DM, we consider speeds above the halo cutoff and treat the incident flux as an input without specifying a production mechanism. We examine monoenergetic and Gaussian energy distributions and scan the mass and splitting
while fixing the coupling. The higher incident speeds allow substantially lighter DM than in the halo case. A benchmark at $v_0=0.005c$ and $\Phi_0=10^{-2}\,\mathrm{cm^{-2}\,s^{-1}}$ has $m\simeq82.87\,\mathrm{GeV}$ and
$\delta\simeq0.626\,\mathrm{MeV}$. It predicts approximately one accepted event, with most of the signal in the middle bin. In the Gaussian case, broader energy distributions allow fitted DM masses down to a few GeV.

This paper is organized as follows. In Sec.~\ref{Sec2}, we introduce the interaction and kinematics for both non-relativistic and boosted DM. Sec.~\ref{Sec3} describes the event-rate calculation and statistical analysis.
In Sec.~\ref{Sec4}, we present benchmarks and parameter regions for both cases. We conclude in Sec.~\ref{Sec5}.
\section{Inelastic DM}\label{Sec2}
\subsection{Inelastic scattering and parameters}
We consider the photon-mediated inelastic scattering
\begin{equation}
 \chi_1(p)+A(k)
 \longrightarrow
 \chi_2(p')+A(k'),
 \label{eq:reaction}
\end{equation}
where $\chi_1=\widetilde a$ is the axino DM and $\chi_2\simeq\widetilde B$ is a heavier bino-like neutralino. The nucleus $A$ has mass $M_A$ and remains in
its ground state. The mass hierarchy is parametrized as 
\begin{equation}
 m_1=m,\qquad m_2=m+\delta,\qquad \delta>0,
 \label{eq:masses}
\end{equation}
so that Eq.~\eqref{eq:reaction} describes an endothermic transition. The relevant interaction originates from the effective axino--bino--hypercharge operator~\cite{Covi:1999ty,Covi:2001nw},
\begin{equation}
 \mathcal L=
 i\frac{\alpha_Y C_{aYY}}{16\pi F_a}
 \overline{\widetilde a}\gamma_5
 [\gamma^\mu,\gamma^\nu]\widetilde B_Y B_{\mu\nu},
 \label{eq:lagY}
\end{equation}
where $\alpha_Y$ is the hypercharge fine-structure constant and $C_{aYY}$ is the anomaly coefficient. The fields $\widetilde B_Y$ and $B_{\mu\nu}$ denote the hypercharge gaugino and field strength. We use $F_a=f_a/N_{\rm DW}$, where $f_a$ is the PQ symmetry breaking scale and $N_{\rm DW}$ is the domain wall number. After electroweak symmetry breaking, $B_{\mu\nu}=c_WF_{\mu\nu}-s_WZ_{\mu\nu}$, the photon interaction becomes
\begin{equation}
 \mathcal L_\gamma =
 \frac{g_\gamma}{2}
 \overline{\chi_1}\sigma^{\mu\nu}\gamma_5
 \chi_2 F_{\mu\nu},\qquad
 g_\gamma=
 \frac{\alpha C_{aYY}N_{11}}{4\pi c_WF_a},
 \label{eq:laggamma}
\end{equation}
where $\alpha=e^2/(4\pi)$, $c_W=\cos\theta_W$, and $N_{11}$ is the bino component of $\chi_2$. We adopt the electric-transition structure and define the photon vertex as
\begin{equation}
 \Gamma^\mu(q)=
 g_\gamma\sigma^{\mu\nu}q_\nu\gamma_5,
 \label{eq:vertex}
\end{equation}
where $q$ is the momentum carried by the exchanged photon. The scattering cross section depends on $|g_\gamma|^2$.
The interaction strength can equivalently be expressed through
\begin{equation}
 F_{a,\mathrm{eff}}
 \equiv\frac{F_a}{|C_{aYY}N_{11}|},
 \qquad
 |g_\gamma|=
 \frac{\alpha}{4\pi c_WF_{a,\mathrm{eff}}}.
 \label{eq:faeff}
\end{equation}
For $C_{aYY}=N_{11}=1$, one has
$F_{a,\mathrm{eff}}=F_a$.
The independent scattering parameters are therefore $m$, $\delta$, and $g_\gamma$.  We treat the fermion masses and effective interaction scale as independent inputs and do not require
the associated axion to be the QCD axion.
\subsection{Non-relativistic scattering}
For a nucleus initially at rest, let $E_R$ denote the nuclear recoil energy and $v$ the incident DM speed in the laboratory frame. In the non-relativistic limit, the minimum speed required to
produce a recoil $E_R$ is
\begin{equation}
 v_{\min}^A(E_R)=
 \frac{M_AE_R/\mu_A+\delta}{\sqrt{2M_AE_R}},
 \qquad
 \mu_A=\frac{mM_A}{m+M_A},
 \label{eq:vmin}
\end{equation}
where $\mu_A$ is the DM--nucleus reduced mass. Minimizing $v_{\min}^A(E_R)$ gives the threshold,
\begin{equation}
 v_{\rm th}^A=\sqrt{\frac{2\delta}{\mu_A}}.
 \label{eq:vthr}
\end{equation}
Consequently, for an incident DM with maximum laboratory speed $v_{\max}$, kinematics requires
\begin{equation}
 \delta\leq\frac{\mu_Av_{\max}^2}{2}.
 \label{eq:halo_condition}
\end{equation}
At a fixed speed $v>v_{\rm th}^A$, the recoil energy lies within $E_R^-(v)\leq E_R\leq E_R^+(v)$, where the lower and upper bounds are
\begin{equation}
 E_R^\pm(v)=
 \frac{\mu_A^2v^2}{2M_A}
 \left(1\pm\sqrt{1-\frac{2\delta}{\mu_Av^2}}\right)^2.
 \label{eq:endpointsNR}
\end{equation}
This interval reflects the allowed scattering angles. At threshold, the bounds coincide at $E_R=\mu_A\delta/M_A$, and the available phase space vanishes. 

At leading non-relativistic order (See Appendix~\ref{AppendixA} for the details), the spin-averaged differential cross section for the coherent nuclear electric-charge is
\begin{equation}
 \frac{\mathrm d\sigma_A^C}{\mathrm dE_R}
 \simeq
 \frac{\alpha Z^2|g_\gamma|^2}{v^2E_R}
 F_{C,A}^2(Q^2)\,\Theta_A,
 \label{eq:NR_cross}
\end{equation}
where the superscript $C$ denotes the nuclear charge contribution and $Z$ is the nuclear atomic number. The momentum transfer is defined by
\begin{equation}
 q=p-p'=k'-k,\qquad Q^2=-q^2=2M_AE_R.
 \label{eq:transfer}
\end{equation}
The factor $\Theta_A=\Theta[v-v_{\min}^A(E_R)]$ is the Heaviside step function.
At this order, the mass splitting enters the cross section through $v_{\min}^A(E_R)$ in $\Theta_A$. The cross section exhibits a coherent charge enhancement $Z^2$ and is proportional to $|g_\gamma|^2$. The explicit $1/E_R$ dependence originates from
photon exchange and the momentum-dependent transition vertex.
The nuclear charge form factor $F_{C,A}(Q^2)$ accounts for finite nuclear size. We adopt the Helm parametrization~\cite{Helm:1956zz},
\begin{equation}
 F_{C,A}(Q^2)=
 \frac{3j_1(QR_{1,A})}{QR_{1,A}}
 \exp\left(-\frac{Q^2s_A^2}{2}\right),
 \label{eq:helm}
\end{equation}
where $j_1$ is the spherical Bessel function of order one, $R_{1,A}$ is the effective nuclear radius, and $s_A$ characterizes the surface smearing. The form factor is normalized to $F_{C,A}(0)=1$. For natural xenon, the cross section, form factor, and kinematic limits are evaluated separately for each isotope, and the resulting event rates contributions are combined with the corresponding isotopic abundances.
\subsection{Boosted DM scattering}

For boosted DM, we use the exact two-body kinematics of Eq.~\eqref{eq:reaction}, with $m_1=m$ and $m_2=m+\delta$.
In the laboratory frame, the initial nucleus is at rest with $k=(M_A,\mathbf 0)$. The incoming and outgoing DM energies are $E_1=p^0$ and $E_2=p'^0$, respectively. The Mandelstam invariant $s$ is
\begin{equation}
 s=(p+k)^2=m_1^2+M_A^2+2M_AE_1.
 \label{eq:s_exact}
\end{equation}
Here $E_1$ denotes the total incident energy, including the rest mass. The final nuclear energy is
\begin{equation}
 E_{A'}=k'^0
 =\frac{2p\cdot p'+2M_A^2-m_1^2-m_2^2}{2M_A}, 
\end{equation}
and the nuclear recoil energy therefore satisfies
\begin{equation}
 E_R=E_{A'}-M_A
 =\frac{2p\cdot p'-m_1^2-m_2^2}{2M_A}.
 \label{eq:E_nr}
\end{equation}
In the center-of-mass frame, Eq.~\eqref{eq:E_nr} becomes
\begin{equation}
 E_R=
 \frac{2E_1^*E_2^*
       -2p_1^*p_2^*\cos\theta^*
       -m_1^2-m_2^2}{2M_A},
 \label{eq:recoil_cm}
\end{equation}
where $\theta^*$ is the scattering angle, with a star denotes a center-of-mass quantity. The DM energies and momentum magnitudes are
\begin{equation}
\begin{aligned}
 E_1^*&=\frac{s+m_1^2-M_A^2}{2\sqrt{s}},
 & E_2^*&=\frac{s+m_2^2-M_A^2}{2\sqrt{s}},\\
 p_1^*&=\frac{\sqrt{\lambda(s,m_1^2,M_A^2)}}{2\sqrt{s}},
 & p_2^*&=\frac{\sqrt{\lambda(s,m_2^2,M_A^2)}}{2\sqrt{s}},
\end{aligned}
\end{equation}
where
\begin{equation}
 \lambda(a,b,c)=a^2+b^2+c^2-2ab-2bc-2ca,
\end{equation}
is the K\"all\'en function. The angular range $-1\leq\cos\theta^*\leq1$ determines the exact recoil endpoints,
\begin{equation}
 E_{R,\mathrm{exact}}^\pm(s)=
 \frac{2E_1^*E_2^*\pm2p_1^*p_2^*-m_1^2-m_2^2}
 {2M_A}.
 \label{eq:endpoints_exact}
\end{equation}
The plus and minus denote the upper and lower endpoints, respectively. For the endothermic transition, the condition
$s\geq(m_2+M_A)^2$ gives
\begin{equation}
 E_1\geq E_{1,\mathrm{th}}^A
 =m_2+\frac{m_2^2-m_1^2}{2M_A}.
 \label{eq:E_thres}
\end{equation}
The corresponding minimum incident kinetic energy is
\begin{equation}
 K_{\mathrm{th}}^A
 \equiv E_{1,\mathrm{th}}^A-m_1
 =\delta\left(1+\frac{m}{M_A}\right)
  +\frac{\delta^2}{2M_A}.
 \label{eq:kinetic_threshold_exact}
\end{equation}
At threshold, $p_2^*=0$ and the recoil endpoints coincide at
\begin{equation}
 E_{R,\mathrm{th}}^A
 =\frac{m_2^2-m_1^2}{2(M_A+m_2)}
 =\frac{\delta(2m+\delta)}{2(M_A+m+\delta)}.
 \label{eq:recoil_threshold_exact}
\end{equation}

The recoil interval then collapses and the available phase space vanishes. At leading non-relativistic order,
Eq.~\eqref{eq:endpoints_exact} reduces to Eq.~\eqref{eq:endpointsNR}, while the threshold recoil becomes $E_{R,\mathrm{th}}^A\simeq\mu_A\delta/M_A$.

For the coherent nuclear charge contribution, the invariant cross section is
\begin{equation}
\begin{aligned}
 \mathrm d\sigma_A^C={}&
 \frac{\overline{|\mathcal M_A|^2}}
 {4\sqrt{(p\cdot k)^2-m_1^2M_A^2}}\,
 (2\pi)^4\delta^{(4)}(p+k-p'-k')
 \\
 &\times
 \frac{\mathrm d^3\mathbf p'}{(2\pi)^3\,2E_2}
 \frac{\mathrm d^3\mathbf k'}{(2\pi)^3\,2E_{A'}}.
\end{aligned}
\end{equation}
Integrating over the remaining phase-space variables yields
\begin{equation}
 \frac{\mathrm d\sigma_A^C}{\mathrm dE_R}
 =
 \frac{M_A}{8\pi\lambda(s,m_1^2,M_A^2)}
 \overline{|\mathcal M_A|^2}\,
 \Theta_A^{\mathrm{exact}},
 \label{eq:cross_exact}
\end{equation}
where $\overline{|\mathcal M_A|^2}$ is the full spin-averaged squared charge amplitude in Eq.~\eqref{eq:invariant_sqm}. The factor $\Theta_A^{\mathrm{exact}}$ equals unity when $E_1\geq E_{1,\mathrm{th}}^A$ and
$E_{R,\mathrm{exact}}^-\leq E_R \leq E_{R,\mathrm{exact}}^+$, and vanishes otherwise. Its leading non-relativistic limit is $\Theta_A$ used in Eq.~\eqref{eq:NR_cross}.
\section{Event rates and LZ statistical analysis}\label{Sec3}
\subsection{Incident distribution and recoil spectrum}
We assume that axinos constitute the local DM, with density $\rho_{\rm in}=0.3\,\mathrm{GeV\,cm^{-3}}$. Using Eq.~\eqref{eq:NR_cross}, the differential recoil rate per unit detector mass is
\begin{equation}
 \frac{\mathrm dR}{\mathrm dE_R}
 =
 \frac{\rho_{\rm in}\alpha Z^2|g_\gamma|^2}{mE_R}
 \sum_A \mathcal N_A
 F_{C,A}^2(2M_AE_R)\,
 \eta\!\left(v_{\min}^A(E_R)\right),
 \label{eq:recoil_rate}
\end{equation}
where $Z=54$ for xenon and $\mathcal N_A$ is the number of nuclei of isotope $A$ per unit detector mass.
For isotopic number fraction $x_A$, $\mathcal N_A=x_A N_{\rm Av}/\overline M_{\rm mol}$, where $N_{\rm Av}$ is Avogadro's constant and $\overline M_{\rm mol}$ is the abundance-weighted mean molar mass of natural xenon. The mean inverse-speed function is
\begin{equation}
 \eta(v_{\min})=
 \int_{v\geq v_{\min}}\mathrm d^3v\,
 \frac{f_{\rm lab}(\mathbf v)}{v},
 \label{eq:inverse_speed}
\end{equation}
where $v=|\mathbf v|$ is the laboratory-frame DM speed and $v_{\min}=v_{\min}^A(E_R)$. The normalized laboratory distribution is $f_{\rm lab}(\mathbf v)=f_{\rm gal}(\mathbf v+\mathbf v_e)$,
where $f_{\rm gal}$ is the Galactic-frame distribution and $\mathbf v_e$ is the laboratory velocity.
We take $v_e=7.3\times10^{-4}c$ and $v_{\rm esc}=1.81\times10^{-3}c$, giving $v_{\max}=v_{\rm esc}+v_e
\simeq761.47\,\mathrm{km\,s^{-1}}$.

\subsection{Detector response}
The reconstructed recoil-energy spectrum is obtained by convolving the true recoil rate in Eq.~(\ref{eq:recoil_rate}) with the detector efficiency and energy response
\begin{equation}
 \frac{\mathrm dN}{\mathrm dE_d}
 =
 \mathcal E
 \int_0^\infty \mathrm dE_R\,
 \epsilon(E_R)\,G(E_d;E_R)\,
 \frac{\mathrm dR}{\mathrm dE_R},
 \label{eq:detected_spectrum}
\end{equation}
where $E_d$ is the reconstructed nuclear recoil energy, $\mathcal E=2.84\,\mathrm{tonne\,yr}$ is the exposure, and $\epsilon(E_R)$ is the detection efficiency. The energy response is approximated by a normalized Gaussian,
\begin{equation}
 G(E_d;E_R)=
 \frac{1}{\sqrt{2\pi}\sigma_E(E_R)}
 \exp\!\left[
 -\frac{(E_d-E_R)^2}{2\sigma_E^2(E_R)}
 \right],
 \label{eq:gaussian_response}
\end{equation}
with the effective resolution
\begin{equation}
 \frac{\sigma_E(E_R)}{\mathrm{keV}}
 =
 \sqrt{
 1.46^2\frac{E_R}{\mathrm{keV}}
 +
 \left(0.093\frac{E_R}{\mathrm{keV}}\right)^2
 }.
 \label{eq:energy_resolution}
\end{equation}
This approximation combines the statistical and systematic energy uncertainties. Following Ref.~\cite{Fan:2026hzw}, we divide the reconstructed energy into three bins with edges
\begin{equation}
 (E_0,E_1,E_2,E_3)
 =
 (70,\,215.5,\,280.5,\,305)\,\mathrm{keV}.
 \label{eq:bin_edges}
\end{equation}
The expected signal count in the $i$-th bin, $E_{i-1}\leq E_d<E_i$, is
\begin{equation}
 N_i=
 \int_{E_{i-1}}^{E_i}\mathrm dE_d\,
 \frac{\mathrm dN}{\mathrm dE_d}.
 \label{eq:bin_counts}
\end{equation}
The convolution in Eq.~\eqref{eq:detected_spectrum} allows events to be reconstructed in a different energy bin or outside the analysis window because of the finite energy resolution.

\subsection{Boosted DM event rates}

For boosted DM scattering, we parameterize the incident flux as
\begin{equation}
 \frac{\mathrm d\Phi}{\mathrm dE_1}
 =
 \Phi_0 f_{\rm in}(E_1;E_c,\sigma_{\rm in}),
 \label{eq:boosted_flux}
\end{equation}
where the constant $\Phi_0$ sets the total incident number flux and depends on the boosted DM production
mechanism. We model the incident energy distribution as a normalized Gaussian~\cite{Su:2023zgr},
\begin{equation}
 f_{\rm in}(E_1;E_c,\sigma_{\rm in})
 =
 \mathcal C_{\rm in}
 \exp\!\left[
 -\frac{(E_1-E_c)^2}{2\sigma_{\rm in}^2}
 \right],
 \qquad E_1\geq m.
 \label{eq:boosted_gaussian}
\end{equation}
Here $E_c$ and $\sigma_{\rm in}$ are the Gaussian center and width, and $\mathcal C_{\rm in}$ is fixed by
$\int_m^\infty f_{\rm in}\,\mathrm dE_1=1$. For monoenergetic incident DM with $E_c>m$, the flux reduces to $\mathrm d\Phi/\mathrm dE_1=\Phi_0\delta_{\rm D}(E_1-E_c)$~\cite{Agashe:2014yua}. Summing over the natural xenon isotopes, the
differential recoil rate per unit detector mass is
\begin{equation}
  \small
 \frac{\mathrm dR^{\rm boosted}}{\mathrm dE_R}
 =
 \Phi_0\sum_A\mathcal N_A
 \int_{E_{1,\mathrm{th}}^A}^{\infty}
 \mathrm dE_1\,
 f_{\rm in}(E_1;E_c,\sigma_{\rm in})
 \frac{\mathrm d\sigma_A^C}{\mathrm dE_R}(E_1,E_R).
 \label{eq:boosted_recoil_rate}
\end{equation}
We use the exact cross section in Eq.~\eqref{eq:cross_exact} and the isotope-dependent threshold in Eq.~\eqref{eq:E_thres}. The factor $\Theta_A^{\mathrm{exact}}$ restricts each contribution to its allowed recoil range. We then apply the detector convolution in Eq.~\eqref{eq:detected_spectrum} using the same efficiency, resolution, and exposure. Integrating the reconstructed spectrum as in Eq.~\eqref{eq:bin_counts} gives the expected
boosted DM counts in the three bins.

\subsection{Statistical Methods}
We adopt a three-bin analysis~\cite{Fan:2026hzw} with assigned counts $(n_1,n_2,n_3)=(0,1,0)$ and compare them with the predicted signal counts $N_i$ calculated from Eq.~\eqref{eq:bin_counts}. Background contributions are neglected. The predicted $N_i$ need not match $n_i$ exactly in each bin.

For independent Poisson counts, the reduced likelihood is
\begin{equation}
 \mathcal L=
 \prod_{i=1}^{3}
 \frac{N_i^{n_i}e^{-N_i}}{n_i!}
 =
 N_2e^{-N_{\rm tot}},
 \label{eq:three_bin_likelihood}
\end{equation}
where $N_{\rm tot}=\sum_iN_i$. This likelihood incorporates both the event in the middle bin and the absence of events in the outer bins. We quantify the fit using the Poisson deviance
\begin{equation}
 \chi_P^2=
 -2\ln\frac{\mathcal L}{\mathcal L_{\rm sat}}
 =
 2\left(N_{\rm tot}-1-\ln N_2\right),
 \label{eq:poisson_deviance}
\end{equation}
where $\mathcal L_{\rm sat}=e^{-1}$ is the saturated-model likelihood obtained at $N_i=n_i$. Smaller $\chi_P^2$ indicates a better fit.

For the non-relativistic scan, we compute the expected counts using the laboratory-frame SHM distribution, with the incident speed $v$ as the integration variable. The integral accounts for the speed-dependent flux and the scattering threshold of each xenon isotope. the signal counts scale as $N_i\propto g_\gamma^2$. Varying $g_\gamma$ therefore changes only the signal normalization. If its allowed range permits $N_{\rm tot}\simeq 1$, profiling over $g_\gamma$ gives
\begin{equation}
 \chi_{P,\mathrm{prof}}^2=-2\ln f_2,
 \qquad
 f_2=\frac{N_2}{N_{\rm tot}}.
 \label{eq:shape_deviance}
\end{equation}
In this case, the fit quality depends only on the middle-bin fraction~\footnote{Generally, the Poisson deviance in Eq.~\eqref{eq:poisson_deviance} can be rewritten as
$\chi_P^2=-2\ln f_2 + 2\left(N_{\rm tot}-1-\ln N_{\rm tot}\right).$
The first term represents the middle-bin fraction.
The second denotes the deviation of $N_{\rm tot}$ from unity.}.
For each $(m,\delta)$, we minimize over the allowed coupling
\begin{equation}
 \Delta\chi^2(m,\delta)=
 \min_{g_\gamma}
 \chi_P^2(m,\delta,g_\gamma)
 -
 \chi_{P,\min}^2,
 \label{eq:delta_chi_squared}
\end{equation}
where $\chi_{P,\min}^2$ is the global minimum within the adopted parameter ranges. 
For boosted DM, we use the same Poisson likelihood and scan $(m,\delta)$ at fixed incident energy $E_c$, energy width $\sigma_{\rm in}$, flux $\Phi_0$, and coupling $g_\gamma$.
We use $\Delta\chi^2=2.28$ and $4.61$ for nominal 68\% and 90\% two-parameter confidence contours. These asymptotic contours are approximate for a single-event sample.
\section{Results}\label{Sec4}
The parameter scan covers
\begin{equation}
\begin{gathered}
400\leq m/\mathrm{GeV}\leq1000,\qquad
100\leq\delta/\mathrm{keV}\leq1000,\\
10^{-10}\leq g_\gamma/(\mathrm{GeV}^{-1})\leq10^{-3}.
\end{gathered}
\label{eq:s4_domain}
\end{equation}
At each $(m,\delta)$, we integrate the recoil rate over the fixed laboratory-frame SHM distribution in Sec.~\ref{Sec3}. The speed integral runs from $10\,\mathrm{km\,s^{-1}}$ to the SHM cutoff at $761.47\,\mathrm{km\,s^{-1}}$, with the kinematically allowed recoil range evaluated separately for each xenon isotope. We adopt the density,
exposure, and detector response specified in Sec.~\ref{Sec3}.
\subsection{Benchmarks and recoil spectra}\label{sec:s4_benchmarks}
Table~\ref{tab:s4_benchmarks} lists the best-fit benchmark, obtained by minimizing the Poisson deviance over $(m,\delta)$ with the fixed SHM and profiling over $g_\gamma$. The best fit has $m=505.92\,\mathrm{GeV}$, $\delta=325.72\,\mathrm{keV}$, and $g_\gamma=10^{-3}\,\mathrm{GeV}^{-1}$ at the upper coupling bound, which is constrained by the current mono-jet LHC search~\cite{Arina:2020mxo}. It predicts approximately one accepted event, with $70.38\%$ in the middle bin and $\chi_P^2=0.70248$. Only \({}^{136}\mathrm{Xe}\) is kinematically accessible.
\begin{table}[tbp!]
 \centering
 \caption{Best-fit benchmark for the fixed SHM. The coupling saturates
 the upper bound in Eq.~\eqref{eq:s4_domain}.}
 \label{tab:s4_benchmarks}
 \begin{tabular}{lc}
 \hline\hline
 Quantity & Best fit \\
 \hline
 $m$ [GeV] & $505.9201$ \\
 $\delta$ [keV] & $325.71521$ \\
 $g_\gamma$ [$\mathrm{GeV}^{-1}$] & $1.00\times10^{-3}$ \\
 Open isotope & \({}^{136}\mathrm{Xe}\) \\
 \hline
 $N_1$ & $0.16667$ \\
 $N_2$ & $0.70047$ \\
 $N_3$ & $0.12809$ \\
 $N_{\rm tot}$ & $0.99524$ \\
 $f_2$ & $0.70382$ \\
 $\chi_P^2$ & $0.702478$ \\
 \hline\hline
 \end{tabular}
\end{table}

\begin{figure}[htbp]
\centering
\includegraphics[width=\columnwidth]{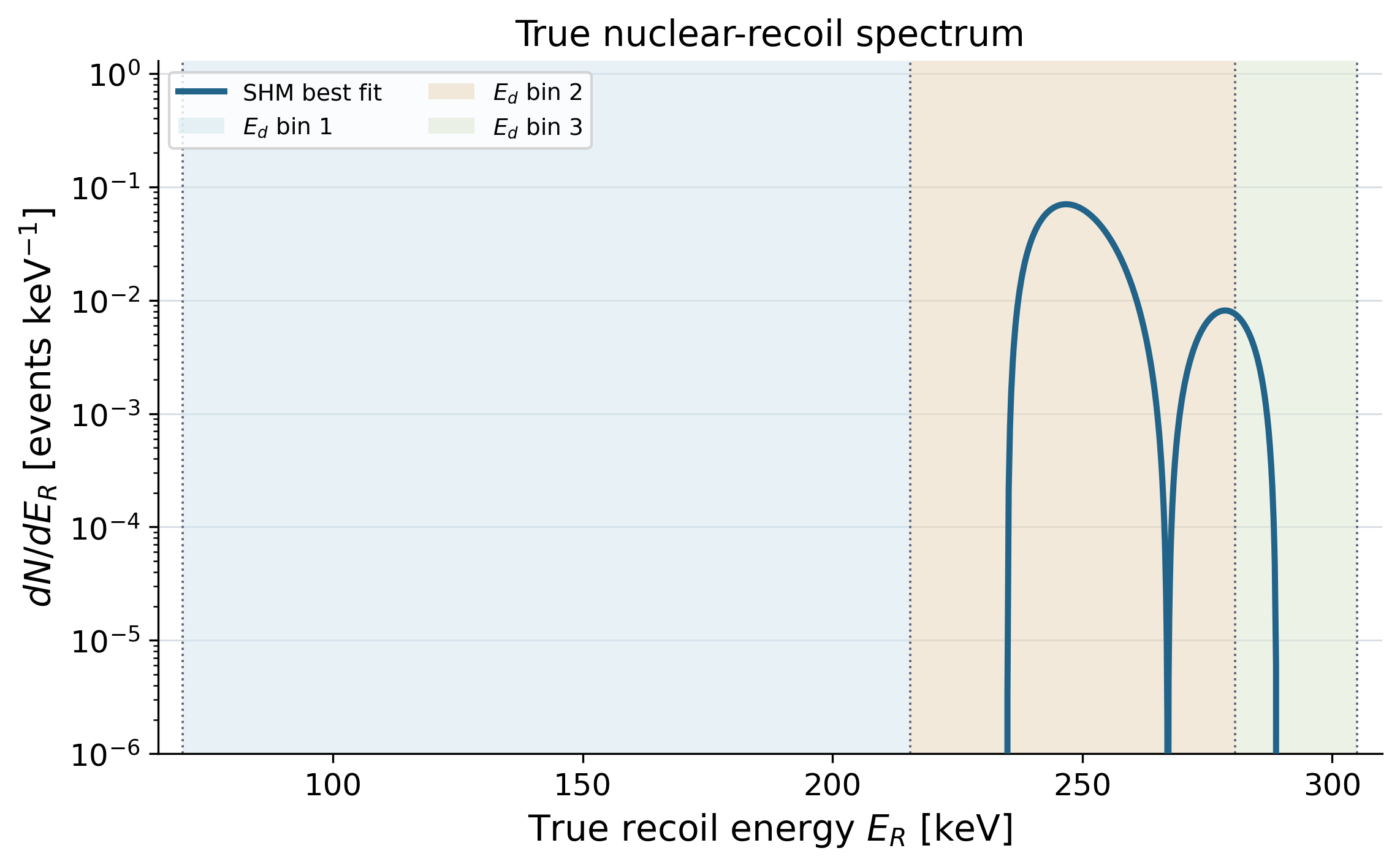}
\caption{
True recoil spectra for the best-fit benchmark. The shaded intervals are three bins.
}
\label{fig:s4_true_recoil}
\end{figure}

Figure~\ref{fig:s4_true_recoil} shows the true recoil spectrum of the best-fit benchmark. We obtain it by integrating the differential scattering rate over the fixed SHM distribution and multiplying by the exposure. Only ${}^{136}\mathrm{Xe}$ contributes at this benchmark.
The kinematically allowed recoil energies range from
$234.97$ to $288.89\,\mathrm{keV}$. The endothermic threshold restricts scattering to the high-speed halo tail, while the nuclear charge form factor produces a minimum near $267\,\mathrm{keV}$ and a smaller peak at higher energy.

\begin{figure}[htbp]
\centering
\includegraphics[width=\columnwidth]{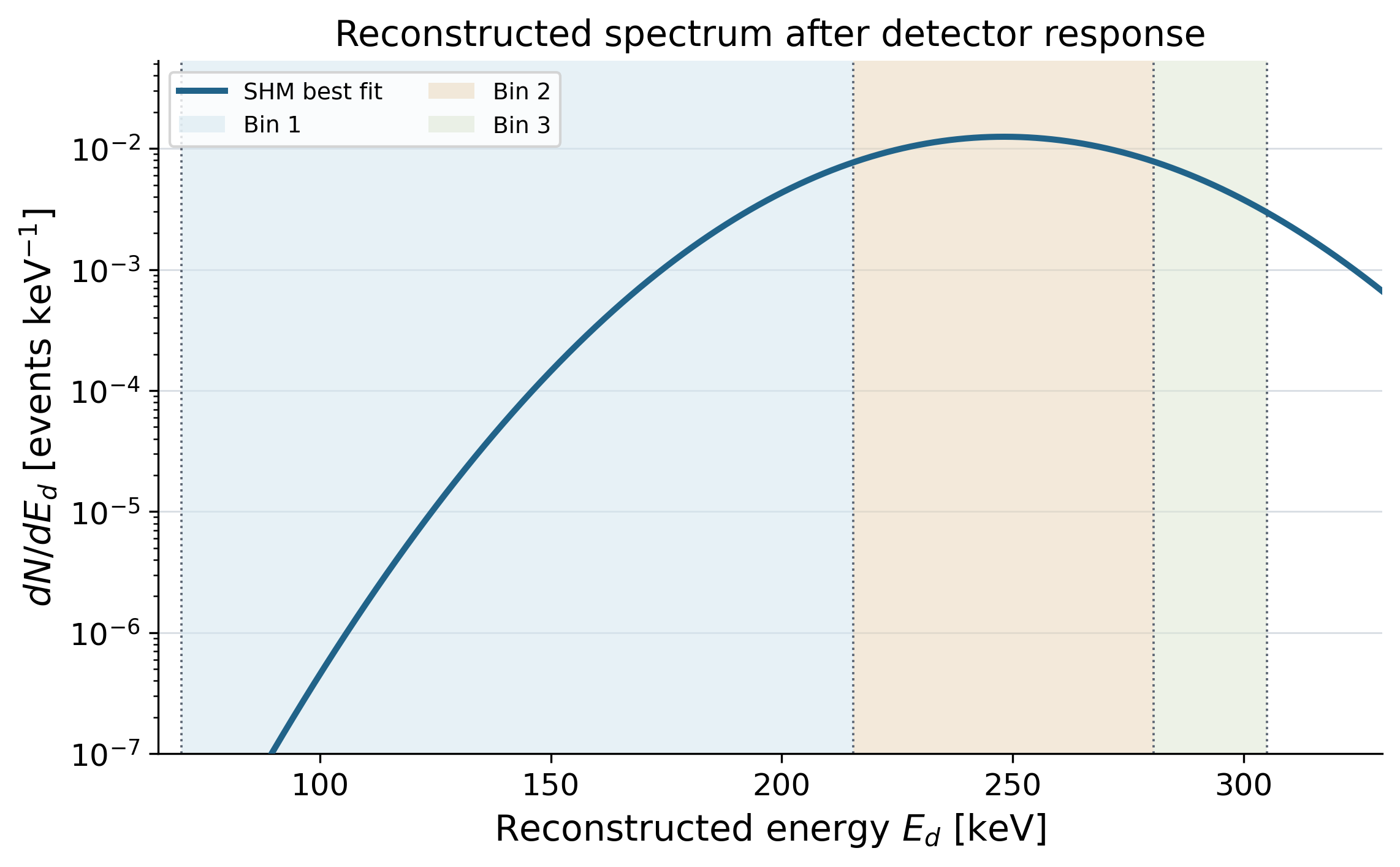}
\caption{
Reconstructed spectra including detector efficiency
and energy resolution.
}
\label{fig:s4_reconstructed}
\end{figure}

Figure~\ref{fig:s4_reconstructed} shows the corresponding reconstructed spectrum. It follows from the efficiency and
energy-resolution convolution in Eq.~\eqref{eq:detected_spectrum}. The finite resolution broadens the spectrum and smooths the form-factor minimum, producing a single maximum near $E_d=248.2\,\mathrm{keV}$. Integrating this spectrum over the three analysis bins gives approximately one accepted event, with $70.38\%$ of the prediction in the middle bin.

\subsection{Binned counts and parameter space}
\label{sec:s4_fit_results}

\begin{figure}[htbp]
\centering
\includegraphics[width=\columnwidth]{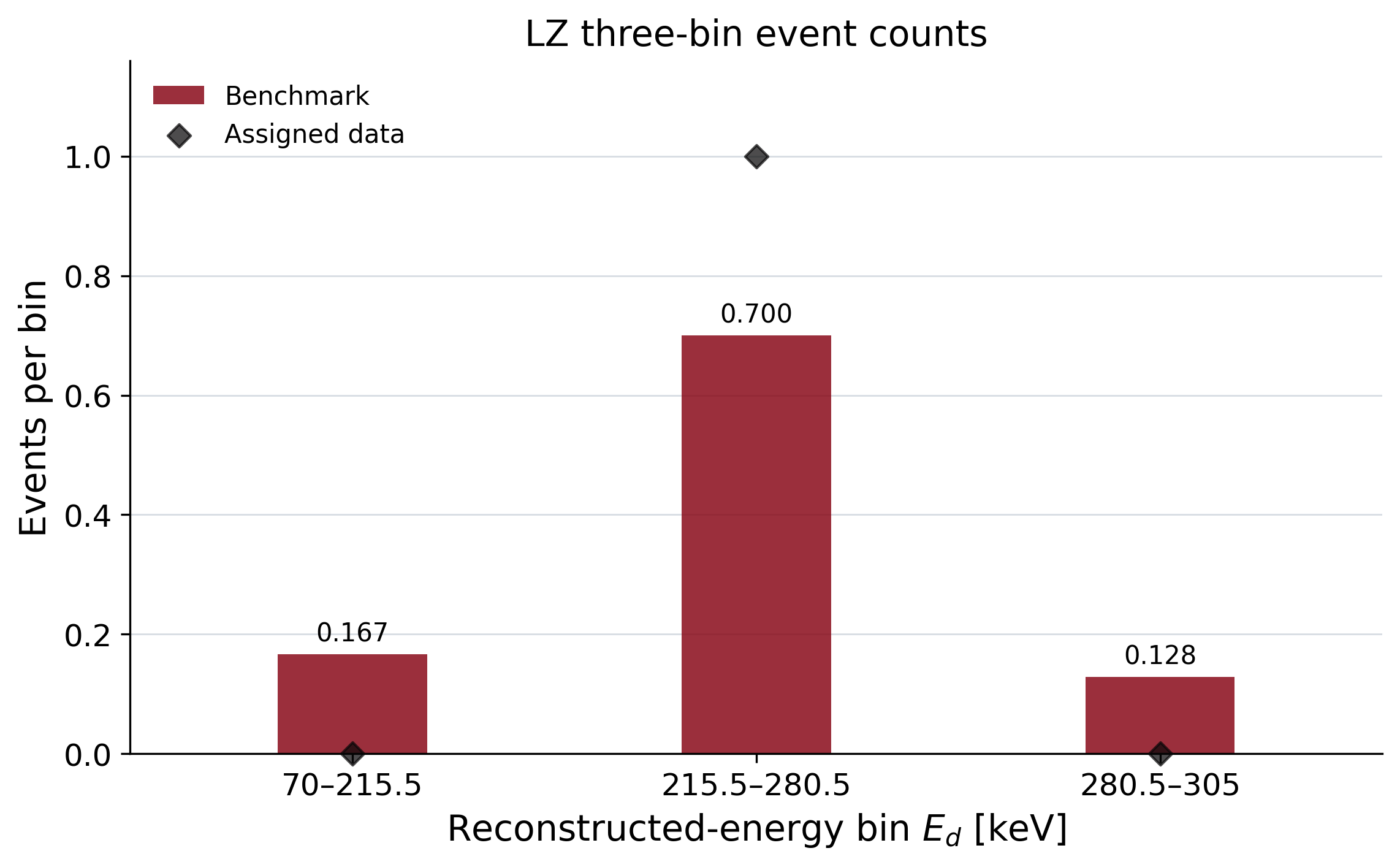}
\caption{
The assigned and expected counts in the three bins.
}
\label{fig:s4_bin_counts}
\end{figure}

Figure~\ref{fig:s4_bin_counts} shows the expected signal counts in the three reconstructed-energy bins, together with the assigned counts $(0,1,0)$. Each bar is obtained by integrating the reconstructed spectrum over the corresponding bin, as defined in Eq.~\eqref{eq:bin_counts}. The benchmark predicts approximately one accepted event in the analysis window, with $70.38\%$ in the middle bin. The smaller contributions to the outer bins arise from the spectral tails and energy reconstruction.

\begin{figure}[htbp]
\centering
\includegraphics[width=\columnwidth]{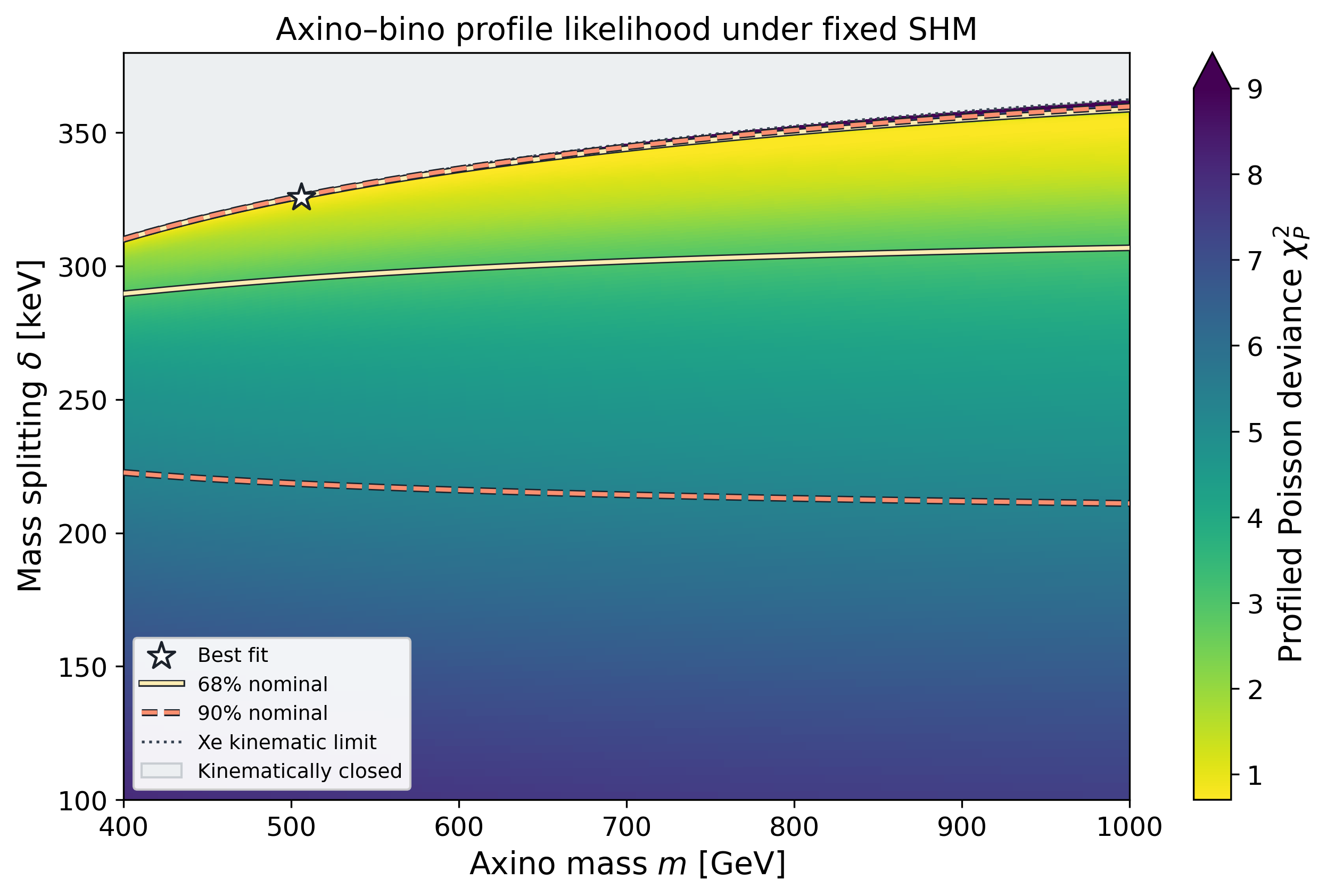}
\caption{
Profiled Poisson deviance in the mass-splitting plane
for the fixed SHM. The color shows the minimum $\chi_P^2$
over the allowed $g_\gamma$ range. Solid and dashed curves
denote the nominal $68\%$ and $90\%$ contours, respectively.
The star marks the best-fit benchmark. The dotted curve
indicates the kinematic limit at $v_{\max}$. Light gray is kinematically forbidden.
}
\label{fig:s4_mass_splitting}
\end{figure}

Figure~\ref{fig:s4_mass_splitting} shows the minimum Poisson deviance in the mass-splitting plane. For each $(m,\delta)$, we calculate the signal counts using the fixed SHM distribution and vary $g_\gamma$ within its allowed range to minimize $\chi_P^2$. The color encodes this minimum, while the nominal $68\%$ and $90\%$ contours are defined by the corresponding $\Delta\chi^2$ thresholds in Eq.~\eqref{eq:delta_chi_squared}. The star marks the best-fit benchmark.

The upper boundary rises with $m$ because heavier DM can supply more energy at the same incident speed, allowing a
larger mass splitting. Equivalently, at fixed $\delta$, increasing $m$ lowers the minimum speed required for scattering and allows more of the halo population to contribute. At each mass, a small $\delta$ favors low-energy recoils and predicts too many events in the first bin. Increasing $\delta$ shifts the signal toward the middle bin and improves the fit. When $\delta$ approaches the upper boundary, however according to Eq.~\eqref{eq:vthr}, only particles near $v_{\max}$ can
scatter, and the upper bound on $g_\gamma$ limits the achievable event rate. Above the dotted line, scattering is kinematically forbidden for all natural xenon isotopes.

\subsection{Boosted DM transition}
\label{sec:s4_boosted}

We consider boosted DM with incident speeds above the halo cutoff, $v_0>v_{\max}$. The sampled speeds are $v_0/c=0.005$, $0.01$, $0.05$, and $0.1$, approximately two to forty times $v_{\max}$. 
Throughout this subsection, we fix $|g_\gamma|=10^{-3}\,\mathrm{GeV}^{-1}$ instead. Two cases of incident DM have been considered. The first is mono-energetic, while the second follows a Gaussian energy distribution with central energy $E_c$ and width $\sigma_{\rm in}$. In the Gaussian case, $v_0$ denotes the speed associated with $E_c$.

\begin{figure}[htbp]
 \centering
 \includegraphics[width=\columnwidth]
 {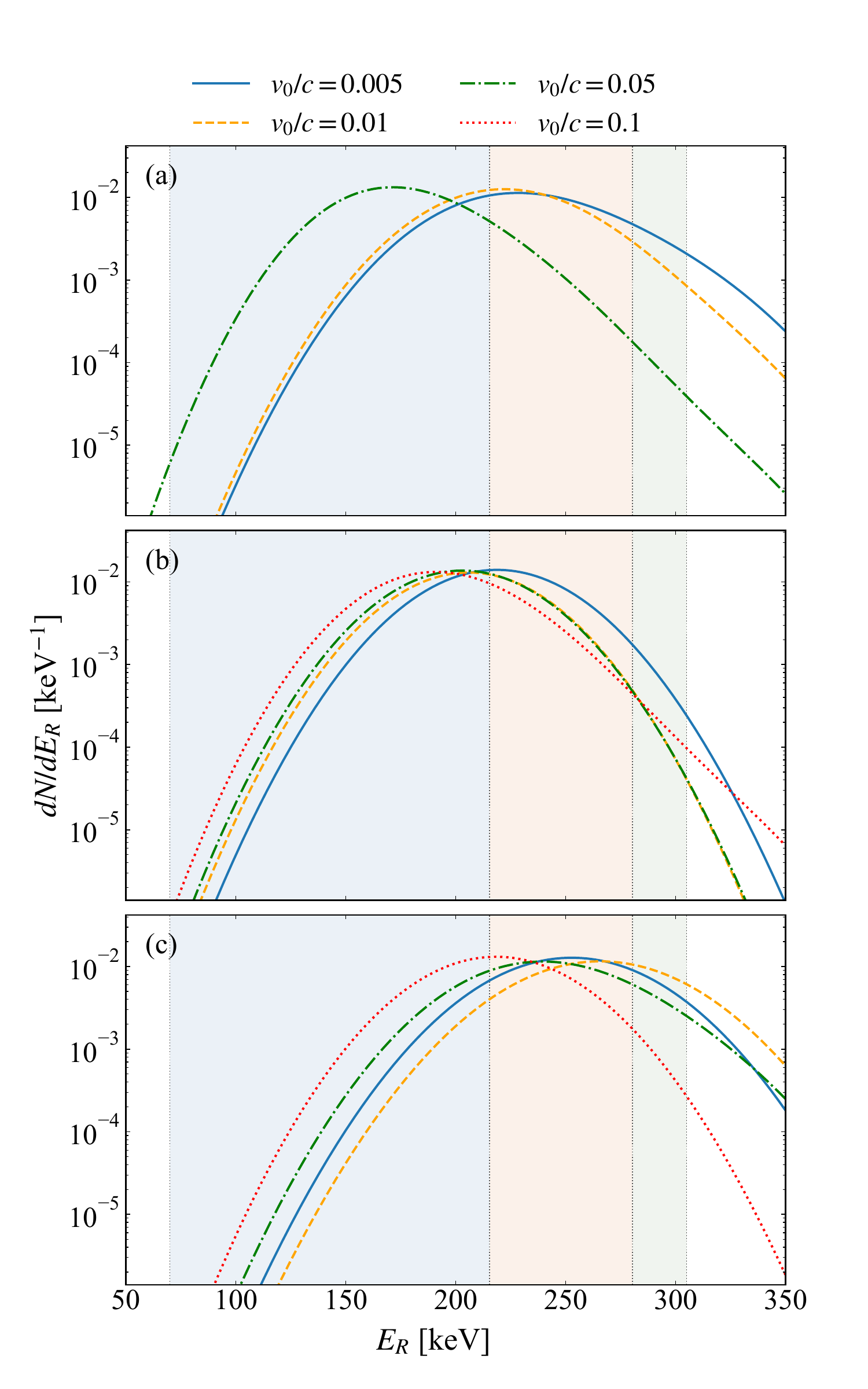}
 \caption{
 Recoil-energy spectra for mono-energetic incident DM.
 Panels (a)--(c) correspond to
 $\Phi_0=10^{-4}$, $10^{-3}$, and
 $10^{-2}\,\mathrm{cm}^{-2}\mathrm{s}^{-1}$,
 respectively.
 The blue solid, orange dashed, green dash-dotted,
 and red dotted curves correspond to
 $v_0/c=0.005$, $0.01$, $0.05$, and $0.1$.
 Each curve uses the parameters listed in
 Table~\ref{tab:bf_fix}.
 The shaded bands indicate the three analysis
 intervals.
 }
 \label{fig_axino_bino}
\end{figure}

Figure~\ref{fig_axino_bino} shows the recoil-energy spectra for mono-energetic DM at three assumed fluxes. For each fixed $(\Phi_0,v_0)$, we scan $(m,\delta)$ and select benchmarks that predict approximately one event and minimize Eq.~\eqref{eq:delta_chi_squared} of boosted case (with $g_\gamma$ omitted and $\min\chi_{P}^2$
replaced by the minimum over $m$ and $\delta$ at fixed
$\Phi_0$ and $v_0$) in the reconstructed three-bin analysis. The corresponding parameters and counts are listed
in Table~\ref{tab:bf_fix}.

\begin{table}[htbp]
 \centering
 \caption{
 Selected benchmark parameters and expected counts
 for mono-energetic incident DM.
 The coupling is fixed at
 $|g_\gamma|=10^{-3}\,\mathrm{GeV}^{-1}$.
 The smallest listed deviance is shown in bold.
 A dash indicates that no benchmark is reported
 for that configuration.
 }
 \label{tab:bf_fix}
 \begin{tabular}{ccccccc}
 \toprule
 $\Phi_0$ [$\mathrm{cm}^{-2}\mathrm{s}^{-1}$]
 & $v_0/c$
 & $m$ [GeV]
 & $\delta$ [MeV]
 & $\chi_P^2$
 & $N_{\rm tot}$
 & $N_2$ \\
 \midrule
 $10^{-2}$ & 0.005 & 82.87 & 0.6258
 & {\color{red}0.805} & {\color{red}1.06} & {\color{red}0.714} \\
 & 0.01 & 32.61 & 1.296
 & 1.047 & 1.00 & 0.598 \\
 & 0.05 & 5.405 & 6.434
 & 0.923 & 1.02 & 0.643 \\
 & 0.1 & 2.365 & 11.68
 & 1.311 & 1.01 & 0.525 \\
 \midrule
 $10^{-3}$ & 0.005 & 75.85 & 0.5926
 & 1.278 & 1.05 & 0.555 \\
 & 0.01 & 28.03 & 1.147
 & 2.015 & 0.970 & 0.346 \\
 & 0.05 & 4.673 & 5.632
 & 2.194 & 1.01 & 0.341 \\
 & 0.1 & 2.847 & 13.1
 & 2.932 & 1.00 & 0.233 \\
 \midrule
 $10^{-4}$ & 0.005 & 98.74 & 0.6713
 & 1.150 & 1.05 & 0.593 \\
 & 0.01 & 35.37 & 1.334
 & 1.209 & 1.03 & 0.567 \\
 & 0.05 & 5.973 & 6.411
 & 4.470 & 1.01 & 0.108 \\
 & 0.1 & \multicolumn{5}{c}{---} \\
 \bottomrule
 \end{tabular}
\end{table}

In Figure~\ref{fig_axino_bino}, the peaks move toward
the middle bin as the assumed flux increases
from panel (a) to panel (c).
This shift arises because the mass and splitting
are refitted at each flux.
At fixed $(m,\delta,v_0)$, changing the flux only
rescales the spectrum.
At the lowest flux, the high-speed case $v_0=0.1c$
yields too few events, therefore no benchmark is listed for this case
in Table~\ref{tab:bf_fix}.
Spectra with larger deviance tend to peak at lower
recoil energies and predict fewer middle-bin events,
even when the total count is close to one.
At each $v_0$, the fitted $m$ and $\delta$ must
satisfy the incident-energy threshold in
Eq.~\eqref{eq:E_thres}.
Near this threshold, a smaller mass gives a lower
recoil energy at fixed $\delta$.
The best listed fit occurs at
$v_0=0.005c\simeq2v_{\max}$ and
$\Phi_0=10^{-2}\,\mathrm{cm}^{-2}\mathrm{s}^{-1}$,
predicting $N_{\rm tot}=1.06$ and $N_2=0.714$.

Table~\ref{tab:benchmark} lists the adopted lower
mass limits $m_{\min}$ for different incident
speeds and Gaussian energy widths.
For mono-energetic DM, $m_{\min}$ decreases from
$66.85$ to $2.198\,\mathrm{GeV}$ as $v_0$ increases
from $0.005c$ to $0.1c$.
This speed dependence reflects the incident-energy
threshold for ${}^{136}\mathrm{Xe}$ in
Eq.~\eqref{eq:E_thres}.
At positive splitting, lower speeds require
larger masses to cross this threshold.
For $\sigma_{\rm in}/E_c=10^{-3}$, however,
the limits remain between $1.731$ and
$2.859\,\mathrm{GeV}$ over the same sampled speeds,
showing a much weaker dependence on the central speed. This behavior arises because particles in the high-energy tail can scatter even when $E_c$ lies below threshold.
A broader distribution can therefore provide
appreciable rates at smaller masses.

\begin{table}[htbp]
 \centering
 \caption{
 Adopted lower limits $m_{\min}$ of the mass scans,
 in GeV, for mono-energetic and Gaussian incident DM.
 The Gaussian columns are labeled by
 $\sigma_{\rm in}/E_c$.
 All scans extend from $m_{\min}$ to
 $400\,\mathrm{GeV}$.
 }
 \label{tab:benchmark}
 \setlength{\tabcolsep}{4pt}
 \begin{tabular}{ccccc}
 \toprule
 & & \multicolumn{3}{c}{Gaussian} \\
 \cmidrule(lr){3-5}
 $v_0/c$ & Mono-energetic
 & $10^{-7}$ & $10^{-5}$ & $10^{-3}$ \\
 \midrule
 0.005 & 66.85 & 65.64 & 29.07 & 2.859 \\
 0.01  & 26.32 & 26.22 & 19.92 & 2.841 \\
 0.05  & 4.493 & 4.492 & 4.438 & 2.396 \\
 0.1   & 2.198 & 2.198 & 2.191 & 1.731 \\
 \bottomrule
 \end{tabular}
\end{table}

\begin{figure}[htbp]
 \centering
 \includegraphics[width=\columnwidth]{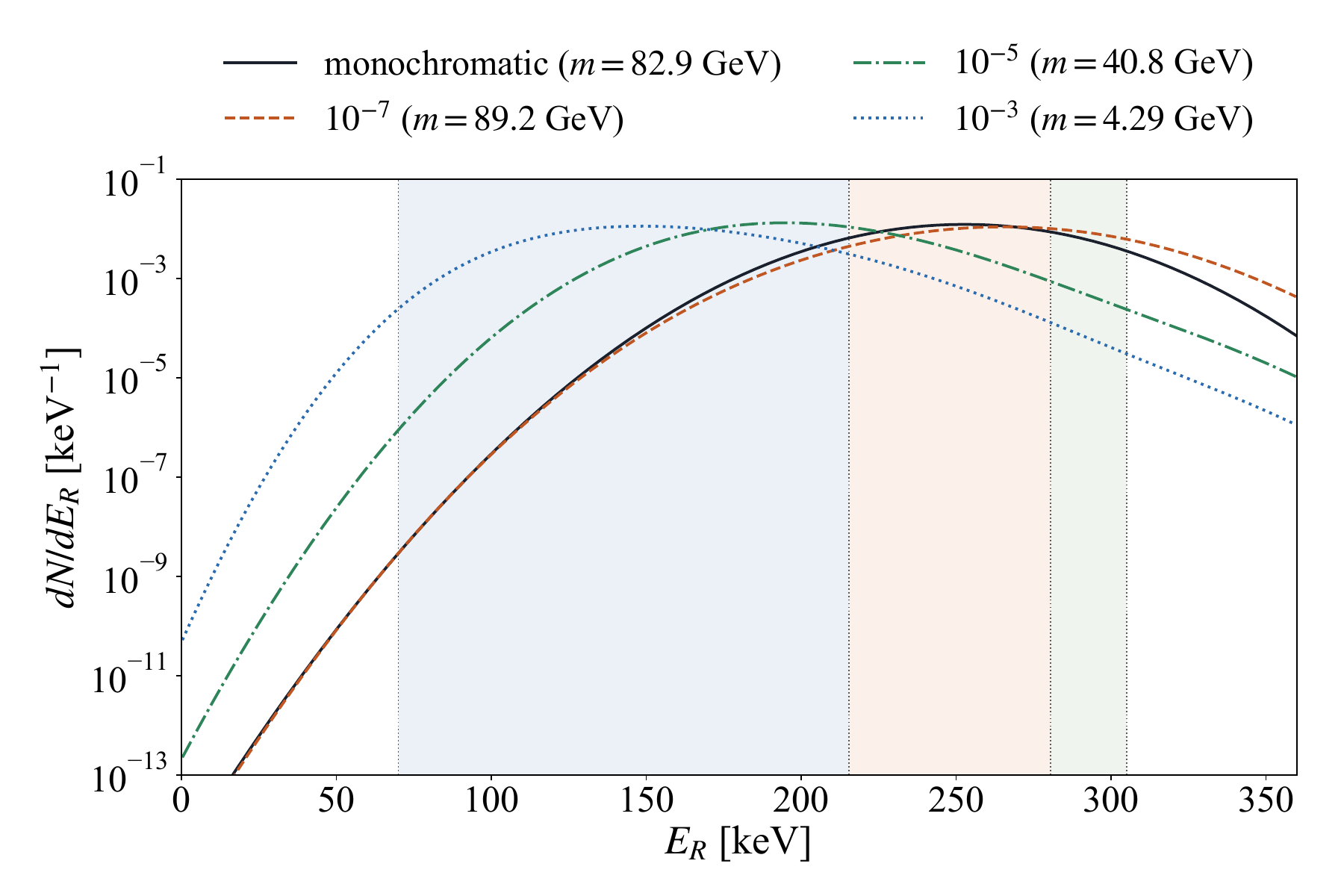}
 \caption{
 Recoil-energy spectra for different incident
 energy distributions at
 $\Phi_0=10^{-2}\,\mathrm{cm}^{-2}\mathrm{s}^{-1}$
 and $v_0=0.005c$. The solid curve denotes the mono-energetic case. Other line styles indicate Gaussian incident energy distributions with $\sigma_{\rm in}/E_c=10^{-7}$, $10^{-5}$, and $10^{-3}$.
 }
 \label{fig_sigmaE}
\end{figure}

Figure~\ref{fig_sigmaE} compares the event spectra
for different incident energy distributions.
Each benchmark is selected by minimizing Eq.~\eqref{eq:delta_chi_squared} while keeping the expected
total count close to one (with $g_\gamma$ omitted and $\min\chi_{P}^2$
replaced by the minimum over $m$ and $\delta$ at fixed
$E_c$, $\sigma_{\rm in}$, and $\Phi_0$).
The narrow Gaussian case remains close to the
mono-energetic one, whereas broader distributions
yield smaller fitted masses and peaks at lower
recoil energies. Specifically, the mono-energetic case has $m=82.9\,\mathrm{GeV}$
and a peak near $E_R=250\,\mathrm{keV}$.
The narrow Gaussian case with
$\sigma_{\rm in}/E_c=10^{-7}$ gives
$m=89.2\,\mathrm{GeV}$ and a peak near
$265\,\mathrm{keV}$.
Increasing the relative width to $10^{-5}$ and
$10^{-3}$ reduces the fitted masses to
$40.8$ and $4.29\,\mathrm{GeV}$, respectively.
The corresponding peaks move to no more than
$200$ and $150\,\mathrm{keV}$ respectively.

\section{conclusion}\label{Sec5}
The LZ experiment extended its nuclear-recoil search to approximately $270\,\mathrm{keV}$ and reported a candidate at $248\pm23\,\mathrm{(stat)}\pm23\,\mathrm{(sys)}\,\mathrm{keV}$ in an exposure of $2.84\,\mathrm{tonne\,yr}$. 
We have investigated the photon-mediated endothermic axino-bino transition $\widetilde a A\to\widetilde B A$
to explain this high-energy event. We calculated the signals for non-relativistic halo DM and boosted DM using the corresponding scattering cross sections and kinematics. The analysis includes natural xenon isotopes, nuclear charge form factors, detector efficiency, and energy resolution.  We compared the reconstructed event counts with the assigned LZ counts using a simplified three-bin Poisson likelihood. The benchmarks are selected by minimizing the corresponding deviance and predict approximately one event in total, with the signal dominated by the middle bin.

For halo DM, we scanned the DM mass, positive mass splitting, and coupling. We integrated the rate over a fixed
laboratory-frame SHM distribution. The best-fit benchmark has $m\simeq506\,\mathrm{GeV}$, $\delta\simeq326\,\mathrm{keV}$, and $g_\gamma=10^{-3}\,\mathrm{GeV}^{-1}$, at the adopted upper coupling bound. It predicts approximately one accepted event, with about $70\%$ in the middle bin. The endothermic threshold suppresses low-energy recoils while restricting scattering to the high-speed halo tail. The fit reaches the upper bound of the coupling to compensate for this small available flux.
For boosted DM, we treated the incident flux as an input and examined mono-energetic and Gaussian energy distributions. At the same fixed coupling, a benchmark with $m\simeq82.87\,\mathrm{GeV}$ and $\delta\simeq0.626\,\mathrm{MeV}$ provides a comparable fit for $v_0=0.005c$ and $\Phi_0=10^{-2}\,\mathrm{cm^{-2}\,s^{-1}}$.
It also predicts approximately one accepted event dominated by the middle bin. 
A Gaussian energy spread allows boosted DM to scatter even when the central energy lies below threshold.
If the energy spread is sufficiently large, the fitted DM mass can decrease to a few GeV.

Both scenarios can fit the LZ candidate within the three-bin analysis, with lighter DM allowed
in the boosted case. Additional recoil data would help distinguish the two interpretations.

\acknowledgments
TL is supported in part by the National Key Research and Development Program of China Grant No. 2020YFC2201504, by the Projects No. 11875062, No. 11947302, No. 12047503, and No. 12275333 supported by the National Natural Science Foundation of China, by the Key Research Program of the Chinese Academy of Sciences, Grant No. XDPB15, by the Scientific Instrument Developing Project of the Chinese Academy of Sciences, Grant No. YJKYYQ20190049, by the International Partnership Program of Chinese Academy of Sciences for Grand Challenges, Grant No. 112311KYSB20210012, and by the Henan Province Outstanding Foreign Scientist Studio Project, No.GZS2025008. This work was supported by the High Performance Computing Platform of Henan Normal University. 

\appendix

\section{Non-relativistic cross section}\label{AppendixA}
\subsection{Conventions}
We consider the endothermic transition
\begin{equation}
 \chi_1(p)+A(k)\longrightarrow\chi_2(p')+A(k'),
 \label{eq:abd_process}
\end{equation}
where $\chi_1=\widetilde a$ is the axino and
$\chi_2\simeq\widetilde B$ is a bino-like neutralino.
Their masses are $m_1=m$ and $m_2=m+\delta$, with
$\delta>0$.
The nucleus has mass $M_A$ and charge $Ze$ and remains
in its ground state.
We use natural units and spinors normalized by
$\overline u_r u_r=2m_r$ for $r=1,2$.

The momentum transfer is $q=p-p'=k'-k$.
In the laboratory frame, the initial nucleus is at rest
and $E_R=k'^0-M_A$.
The exact recoil relations are
\begin{equation}
 \begin{aligned}
 Q^2&\equiv-t=-q^2=2M_AE_R,\\
 |\boldsymbol q|^2&=E_R(E_R+2M_A).
 \end{aligned}
 \label{eq:abd_recoil_relations}
\end{equation}
Thus $|\boldsymbol q|^2\simeq Q^2$ for $E_R\ll M_A$.
The DM energies are $E_1=p^0$ and $E_2=p'^0$,
and the incident velocity is
$\boldsymbol v=\boldsymbol p/E_1$, with
$v=|\boldsymbol v|$.

The transition coupling follows the convention
in Eq.~\eqref{eq:laggamma},
\begin{equation}
 g_\gamma=
 \frac{\alpha C_{aYY}N_{11}}{4\pi c_WF_a},
 \qquad
 |g_\gamma|=
 \frac{\alpha}{4\pi c_WF_{a,\mathrm{eff}}},
 \label{eq:abd_coupling}
\end{equation}
where $\alpha=e^2/(4\pi)$, $c_W=\cos\theta_W$, and
$F_{a,\mathrm{eff}}=F_a/|C_{aYY}N_{11}|$.
The scale $F_a=f_a/N_{\rm DW}$ and the coefficients
$C_{aYY}$ and $N_{11}$ retain their definitions
in the main text.

\subsection{Scattering amplitude and non-relativistic limit}

The physical photon transition vertex is
\begin{equation}
 \Gamma^\mu(q)=
 g_\gamma\sigma^{\mu\nu}q_\nu\gamma_5.
 \label{eq:abd_vertex}
\end{equation}
For a spin-zero nucleus, the charge current with
$e$ factored out is
\begin{equation}
 J_A^\mu=ZF_{C,A}(Q^2)K_A^\mu,
 \qquad K_A=k+k',
 \label{eq:abd_nuclear_current}
\end{equation}
$F_{C,A}(Q^2)$ is the nuclear charge form factor, normalized to
$F_{C,A}(0)=1$.
The one-photon-exchange amplitude is
\begin{equation}
 \mathcal M_A=
 \frac{e g_\gamma ZF_{C,A}(Q^2)}{t}
 K_{A\mu}\,
 \overline u_2(p')\sigma^{\mu\nu}q_\nu\gamma_5u_1(p),
 \label{eq:abd_amplitude}
\end{equation}
up to an overall phase.
The identities $q\cdot K_A=0$ and
$q_\mu\sigma^{\mu\nu}q_\nu=0$ ensure current conservation.

Using the Mandelstam invariants $s=(p+k)^2$ and
$u=(p-k')^2$, the spin-averaged squared amplitude is
\begin{equation}\label{eq:invariant_sqm}
\begin{aligned}
 \overline{|\mathcal M_A|^2}
 &\equiv\frac12\sum_{s_1,s_2}|\mathcal M_A|^2
 \\
 &=\frac{e^2Z^2|g_\gamma|^2F_{C,A}^2(Q^2)}{Q^4}
 \Big\{
 Q^2(s-u)^2
 \\
 &\qquad
 -\delta^2(4M_A^2+Q^2)
 \big[(2m+\delta)^2+Q^2\big]
 \Big\}.
\end{aligned}
\end{equation}
Here $s_1$ and $s_2$ label the initial and final DM spins.
The initial spin average is already included in
the evaluated expression.

For the non-relativistic limit, we expand for
$v\ll1$, $\delta\ll m$, and $E_R\ll M_A$.
We set $m_2\simeq m_1=m$ in leading amplitude factors
while retaining $\delta$ in the kinematic restrictions. The Dirac spinor is
\begin{equation}
 u_r=\sqrt{E_r+m_r}
 \begin{pmatrix}
 \xi_r\\
 \dfrac{\boldsymbol\sigma\cdot\boldsymbol p_r}
 {E_r+m_r}\xi_r
 \end{pmatrix},
 \qquad r=1,2,
 \label{eq:abd_spinor}
\end{equation}
where $\boldsymbol p_1=\boldsymbol p$,
$\boldsymbol p_2=\boldsymbol p'$,
and $\boldsymbol\sigma$ denotes the Pauli matrices.
The two-component spinors satisfy
$\xi_r^\dagger\xi_r=1$.
The temporal component of the transition current is
\begin{align}
 &\overline u_2\sigma^{0i}q_i\gamma_5u_1
 =-i\sqrt{(E_2+m_2)(E_1+m_1)}
 \nonumber\\
 &\quad\times\xi_2^\dagger
 \left[
 \boldsymbol\sigma\cdot\boldsymbol q
 -
 \frac{
 (\boldsymbol\sigma\cdot\boldsymbol p')
 (\boldsymbol\sigma\cdot\boldsymbol q)
 (\boldsymbol\sigma\cdot\boldsymbol p)}
 {(E_2+m_2)(E_1+m_1)}
 \right]\xi_1
 \nonumber\\
 &\quad\simeq
 -2im\,\xi_2^\dagger
 (\boldsymbol\sigma\cdot\boldsymbol q)\xi_1,
 \label{eq:abd_bilinear}
\end{align}
where $i$ is summed over spatial components.
The second term in brackets is suppressed by
two powers of the fermion velocities.
Using $K_A^0=2M_A+E_R\simeq2M_A$ and neglecting
the subleading spatial nuclear current gives
\begin{equation}
 \mathcal M_A^{\rm NR}\simeq
 \frac{4i e g_\gamma ZF_{C,A}(Q^2)mM_A}{Q^2}
 \xi_2^\dagger
 (\boldsymbol\sigma\cdot\boldsymbol q)\xi_1.
 \label{eq:abd_NR_amplitude}
\end{equation}
Using the completeness relation
\begin{equation}
 \frac12\sum_{s_1,s_2}
 \left|
 \xi_2^\dagger
 (\boldsymbol\sigma\cdot\boldsymbol q)\xi_1
 \right|^2=
 \frac12\operatorname{Tr}
 \left[(\boldsymbol\sigma\cdot\boldsymbol q)^2\right]
 =|\boldsymbol q|^2,
 \label{eq:abd_spin_sum}
\end{equation}
and $|\boldsymbol q|^2\simeq Q^2$, the leading
spin-averaged squared amplitude becomes
\begin{equation}
 \overline{|\mathcal M_A^{\rm NR}|^2}
 \simeq
 \frac{16e^2Z^2|g_\gamma|^2m^2M_A^2}{Q^2}
 F_{C,A}^2(Q^2).
 \label{eq:abd_squared_amplitude}
\end{equation}

The exact two-body phase-space result is
\begin{equation}
 \frac{\mathrm d\sigma_A^C}{\mathrm dE_R}
 =
 \frac{M_A}{8\pi\lambda(s,m_1^2,M_A^2)}
 \overline{|\mathcal M_A|^2}\,
 \Theta_A^{\mathrm{exact}},
 \label{eq:abd_recoil_cross_section}
\end{equation}
where $\lambda(a,b,c)=(a-b-c)^2-4bc$ is the
K\"all\'en function.
The factor $\Theta_A^{\mathrm{exact}}$ imposes
the exact energy threshold and recoil bounds.
In the target rest frame,
\begin{equation}
 \lambda(s,m_1^2,M_A^2)
 =4M_A^2(E_1^2-m_1^2)
 \simeq4m^2M_A^2v^2.
 \label{eq:lambda}
\end{equation}

Substituting Eq.~(\ref{eq:abd_squared_amplitude}) and Eq.~(\ref{eq:lambda}) into Eq.~(\ref{eq:abd_recoil_cross_section}) and using $Q^2=2M_AE_R$, we obtain

\begin{equation}
 \boxed{
 \begin{aligned}
 \frac{\mathrm d\sigma_A^C}{\mathrm dE_R}
 &\simeq
 \frac{\alpha Z^2|g_\gamma|^2}{v^2E_R}
 F_{C,A}^2(Q^2)\,\Theta_A
 \\
 &=
 \frac{\alpha^3Z^2}
 {16\pi^2c_W^2F_{a,\mathrm{eff}}^2v^2E_R}
 F_{C,A}^2(Q^2)\,\Theta_A.
 \end{aligned}}
 \label{eq:abd_final}
\end{equation}

\clearpage

\end{document}